\documentclass{article}
\usepackage{authblk}
\usepackage[perpage]{footmisc}

\usepackage{cite}
\usepackage{amsmath,graphicx,url,etoolbox,bm}

\newenvironment{mybullet}{\begin{list}{$\bullet$}
    {\setlength{\topsep}{0mm}\setlength{\itemsep}{0mm}
      \setlength{\parsep}{0.5mm}
      \setlength{\itemindent}{0mm}\setlength{\partopsep}{0mm}
      \setlength{\labelwidth}{15mm}
      \setlength{\leftmargin}{4mm}}}{\end{list}}

\begin{document}
\title{User Blocking Considered Harmful? \\ An Attacker-controllable Side Channel \\ to Identify Social Accounts}



\author[1,2]{Takuya Watanabe\thanks{watanabe.takuya@lab.ntt.co.jp}}
\author[1]{Eitaro Shioji}
\author[1]{Mitsuaki Akiyama}
\author[2]{Keito Sasaoka}
\author[1]{Takeshi Yagi}
\author[2]{Tatsuya Mori\thanks{mori@nsl.cs.waseda.ac.jp}}
\affil[1]{NTT Secure Platform Laboratories}
\affil[2]{Waseda University}

\date{}

\maketitle

\begin{abstract}
This paper presents a practical side-channel attack that identifies the social web service account of a visitor to an attacker's website. Our attack leverages the widely adopted {\em user-blocking} mechanism, abusing its inherent property that certain pages return different web content depending on whether a user is blocked from another user. Our key insight is that an account prepared by an attacker can hold an {\em attacker-controllable} binary state of blocking/non-blocking with respect to an arbitrary user on the same service; provided that the user is logged in to the service, this state can be retrieved as one-bit data through the conventional cross-site timing attack when a user visits the attacker's website. We generalize and refer to such a property as {\em visibility control}, which we consider as the fundamental assumption of our attack. Building on this primitive, we show that an attacker with a set of controlled accounts can gain a complete and flexible control over the data leaked through the side channel. Using this mechanism, we show that it is possible to design and implement a robust, large-scale user identification attack on a wide variety of social web services. To verify the feasibility of our attack, we perform an extensive empirical study using $16$ popular social web services and demonstrate that at least $12$ of these are vulnerable to our attack. Vulnerable services include not only popular social networking sites such as Twitter and Facebook, but also other types of web services that provide social features, e.g., eBay and Xbox Live. We also demonstrate that the attack can achieve nearly $100\%$ accuracy and can finish within a sufficiently short time in a practical setting. We discuss the fundamental principles, practical aspects, and limitations of the attack as well as possible defenses.
\end{abstract}

\section{Introduction}
\label{sec:intro}

The {\em Social web} has become ubiquitous in our daily lives.
It includes not only popular social networking services such as
Facebook and Twitter but also other forms of web services with social
features, e.g., online services for video games such as XBox Live and
online auction/shopping sites such as eBay. 
Social web services facilitate interactions between people with similar
interests.
The widespread adoption of social webs has increased not only the number of
users per service but also the number of services used by
each user.
Smith~\cite{brandwatch_socialmedia} reports that Internet users have an average of five or more social accounts.

Like many other web services, social webs have security and privacy
concerns.
What distinguishes social webs from other web services is that they have 
an intrinsic privacy risk; users are encouraged to share large amounts of
personal/sensitive information on these services, e.g., personal photos,
health information, home addresses, employment status, and sexual preferences.
An attacker can collate various data from social web services to infer
individuals' personal information.
For example, as Minkus et al.~\cite{Minkus_PETS2014} revealed,
an attacker can recover a target's purchase history if s/he knows the
target's eBay account. 
The purchases may include potentially sensitive items, e.g.,
gun-related items or medical tests.
To protect privacy, an eBay user may use a pseudonym for his/her
account name; even in such a case, however, an attacker who can link an
eBay account with an account on Facebook, which encourages users
to disclose their real name, can infer the identity of the actual person
who purchased the sensitive items on eBay.

In this study, we introduce a side-channel attack that identifies the
social account(s) of a website visitor.
The key idea behind our approach is to leverage {\em user blocking},
which is an indispensable mechanism to thwart various types of harassment in
social webs, e.g., trolling, unwanted sexual solicitation, or cyber bullying.
Because user blocking is a generic function commonly adopted by a
wide range of social web services, an attacker can target various
social web services.
In fact, our attack is applicable to at
least the following various social web services:
Ashley Madison, eBay, Facebook, Google+, Instagram,
Medium, Pornhub,  Roblox, Tumblr, Twitter, Xbox Live, and Xvideos.
Because having an account with some of the services included on this list
 could involve privacy-sensitive information,
any account identification can directly lead to privacy risks. 

Our attack leverages the user-blocking mechanism as a means of generating the leaking signals used for the side-channel attack\footnote{More precisely, our side-channel attack is classified as a {\em cross-site timing attack} that will be described in Section~\ref{sec:threat-model}.}. More specifically, we leverage the mechanism's inherent property that certain pages return different web content depending on whether or not a user is blocked from another user. Our key insight is that an account prepared by an attacker can hold an {\em attacker-controllable} binary state of blocking/non-blocking, with respect to an arbitrary user on the service, and this state can be retrieved as one-bit data through cross-site request forgery and a timing side channel when a user visits the attacker's website.
We specially refer to this key action as {\em visibility control} in this paper,
We specifically refer to the property which enables this key action as {\em visibility control} in this paper,
as an attacker is forcing another user to change how they see certain things in the system. Building on this primitive, we show that an attacker can use a set of controlled accounts to construct a {\em controllable} side channel, i.e, leaked data is completely under the attacker's control. Using this mechanism, we show that it is possible to design and implement a robust, large-scale user identification attack mechanism on a wide variety of social web services. We note that the number of accounts required has a theoretically logarithmic relation to the number of users to be targeted, e.g., $20$ attacker-prepared accounts are needed to cover $1$ million users. The novelty of our attack is discussed further in Section~\ref{sec:novelty}.

We note that disabling our side channel, i.e., user blocking, requires careful assessment as it is a crucial function that is widely used on social webs. As we will discuss in Section~\ref{sec:userblocking}, an analysis of 223,487 Twitter users revealed that 3,770 users have blocked more than 1,000 accounts. Our online survey also revealed that 52.3\%/41.4\% of Twitter/Facebook users have responded they have used the blocking mechanism before, and 92.4\%/93.9\% responded there should not be a limit on the number of blocks. These results suggest that neither disabling blocking nor posing a limit on it, is desirable from the viewpoints of the actual usage of the service and users' expectations.
Furthermore, as we show in Section~\ref{sec:attack_extension}, limiting the number of user blocks per account would not be an effective countermeasure owing to our additional technique, {\em user-space partitioning}.

To verify the feasibility of our attack, we performed extensive
empirical studies using 16 existing social web services. 
As mentioned above, we found that 12 of these services are
vulnerable to the attack.
Using $20$ actual accounts, we found
that the attack succeeds with nearly $100\%$ accuracy under a practical setting.

Our contributions can be summarized as follows:
\begin{mybullet}
  \item We demonstrate that the {\em user-blocking} mechanism, which
    is an indispensable function widely adopted in various social web
    services, can be exploited as the leaking signals for a
    side-channel attack that identifies user accounts.
  \item In addition to the side-channel attack, we develop several
    techniques to accurately identify users' accounts. We also reveal
    that this attack is applicable to many currently existing services.
    The attack has a high success rate of nearly 100\%, and is high-speed, taking as short as 4--8 seconds in a preferable setting, or 20–-98 seconds even in a crude environment with a large amount of delay.
  \item We discuss the principles, the practical aspects, and the limitations of this study, as well as some defenses against the attack.
\end{mybullet}

\section{Background: User Blocking}
\label{sec:background}

In this section, we first provide a technical overview of {\em user blocking}, which serves as a side channel used for the user identification attack.  
Next, we demonstrate that simply disabling/limiting this side channel is not a desirable solution against the attack from the viewpoints of actual usage and user expectations.

\subsection{Technical Overview}
\label{sec:userblocking_ov}

{\em User blocking} is a means of blocking communication between two users.
Note that some ``blocking'' mechanisms adopted by social web services
are not {\em user blocking} per se but {\em message blocking}, e.g.,
``muting'' or ``ignoring". While user blocking rejects a person access to
 your account, message blocking filters out all the
messages (or notifications) originating from that person.
Even if a person is blocked with message blocking, this does
not necessarily mean that they do not have access to your
online activities. 
In this paper, we will not focus on message blocking unless
otherwise noted.

\begin{figure}[t]
  \begin{center}
    \includegraphics[width=120mm,clip]{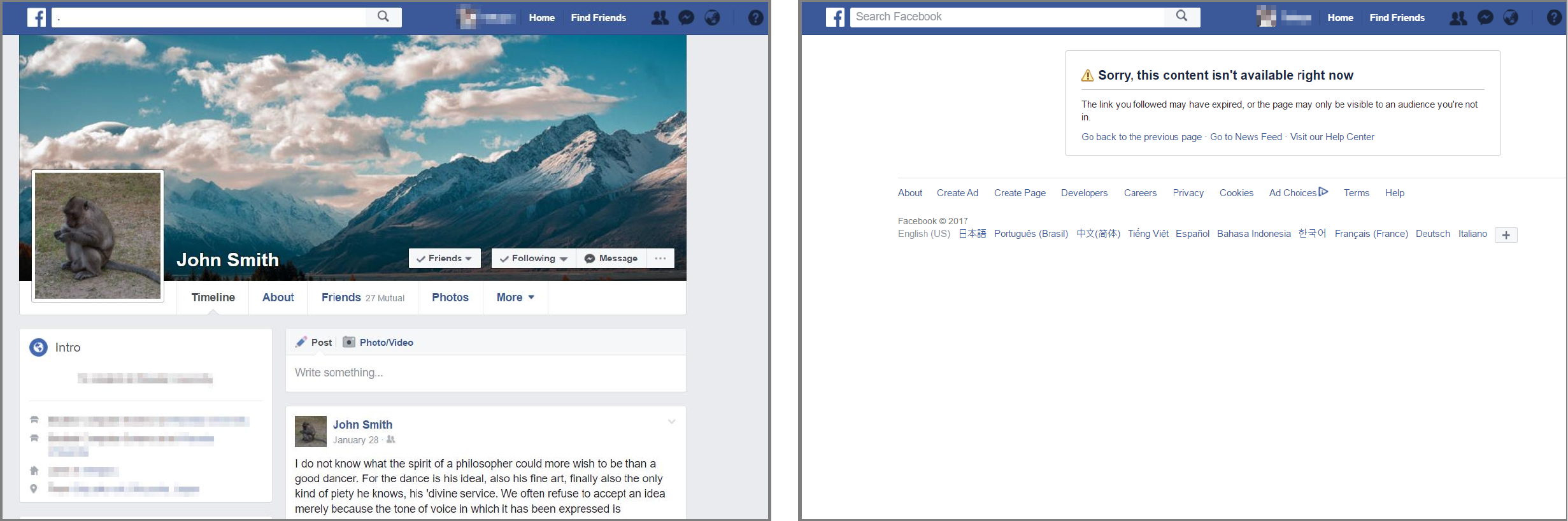}
    \caption{The differences of appearance between non-blocking (left) and blocking (right) pages on Facebook.}
    \vspace{-2mm}
    \label{fig:page_block_unblock}
  \end{center}
\end{figure}

Social web services with {\em user-blocking} mechanisms have intrinsic
web pages that change content depending on the status of the
visitor, i.e., whether or not a visitor is blocked from another person.
A typical example is a user profile data page, which provides information on a person such
as a photograph (icon), a self-introduction, affiliation, recent
posts/updates, etc.
Figure~\ref{fig:page_block_unblock} shows screenshots of
some Facebook profile pages.
In the non-blocked state, the user profile information is fully
available; in the blocked state, these pieces of information are
hidden.
In addition to a user profile page, some social web services provide pages that reflect similar differences. 
A summary of such techniques is presented in Section~\ref{sec:rtt}.

To execute user blocking, a user typically clicks the ``block'' button
set on the profile page of the person to be blocked or enters the account
ID of the person in a text box shown on a dedicated page for
user blocking. Even though official application programming interfaces (APIs) for performing user blocking are not
necessarily provided on all social web services, to the best of our
knowledge no services adopt a special mechanism, such as CAPTCHA,
to prevent automated user-blocking requests.
Therefore, it is currently easy to perform the large-scale
user blocking necessary to implement our user identification attack by using a
script that emulates authentic requests or a headless browser.

\subsection{Usage and Expectations}
\label{sec:userblocking}

\begin{figure}[tbp]
\centering
  \includegraphics[width=120mm,clip]{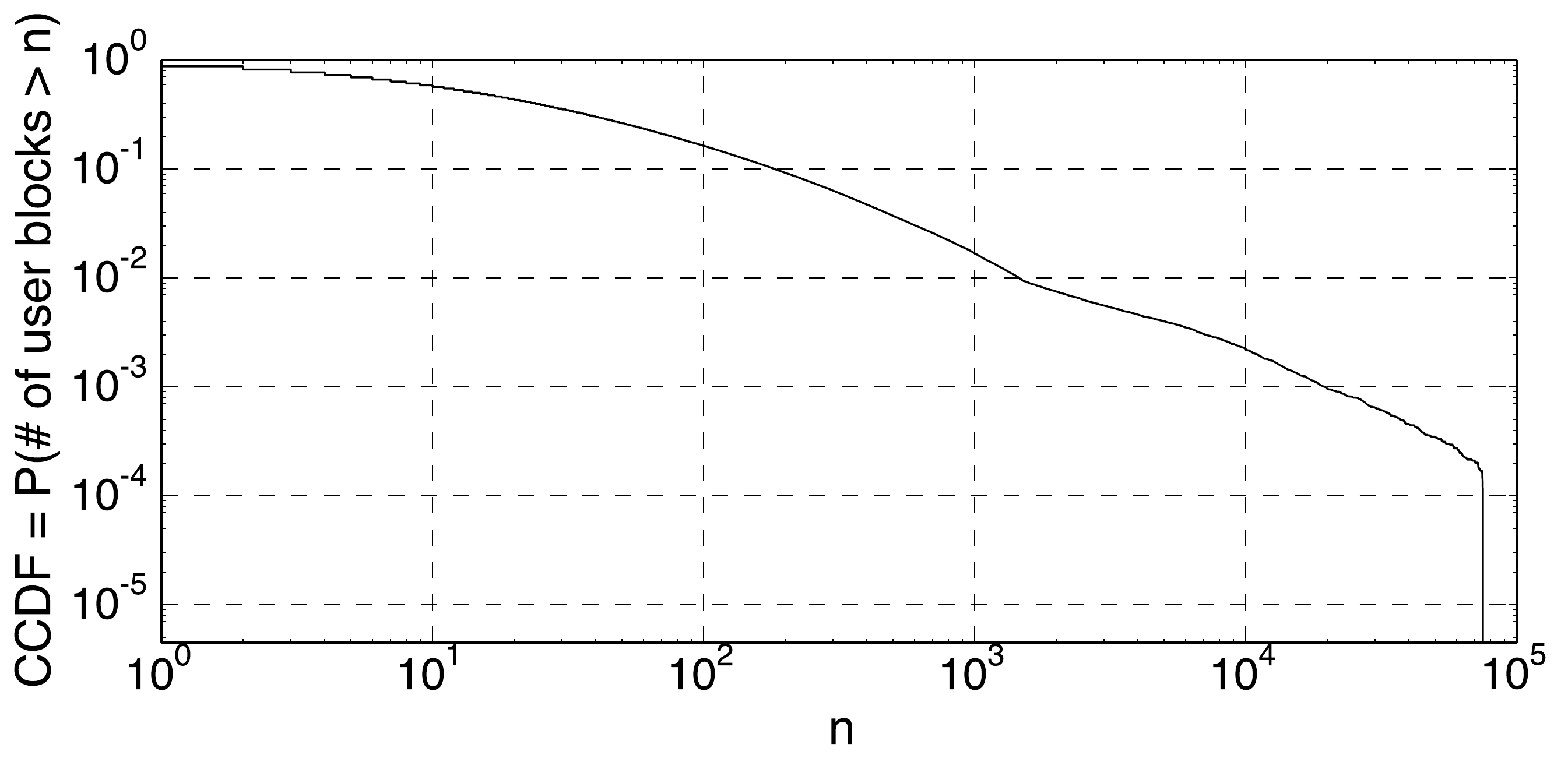}
  \caption{Log-log CCDF of the number of user blocks per account on Twitter. Mean value is $154.21$.}
  \label{fig:twitter_ccdf}
\end{figure}

In this subsection, we discuss how many accounts do people block on social web services and why they do so. 
To answer the ``how many'' question, we first present statistics derived from the data collected at ``Blocked By Me''~\cite{blockedbyme}, a web service that displays a list of users a person has blocked on Twitter\footnote{The data are provided on the courtesy of Gerry Mulvenna~\cite{blockedbyme}. The entire set was anonymized to protect user privacy.}. The data, comprising the number of blocked users for 223,487 unique accounts, were collected from March 2011 to August 2017.
As an individual may have used the web service for several times during the measurement period, we adopt the maximum value of the numbers of blocked users measured for each person.
Figure~\ref{fig:twitter_ccdf} shows the log-log complementary cumulative distribution function (CCDF) of the number of blocked users per account. 
It is seen that the distribution is heavy-tailed, indicating that, although the majority of users blocked a small number of other accounts (median $=$ 15), a non-negligible number of users had to block a large number of other accounts. For instance, 3,770 users blocked more than 1,000 accounts.
Note that the rate-limit of access to Twitter API truncates the number of blocked users at 75,000; thus, users indicated in the figure as having blocked 75,000 users are likely to have actually blocked more.
Besides this upper bound, there were several groups of accounts having the same large number of blocked accounts.
They may be using a shared block list to evade various harassments. As checking the content of such lists is not feasible, some users may have simply cumulatively added new accounts to their block lists. These insights account for the reason why several users have a large number of blocked users. 

\begin{table}[tbp]
  \caption{Demography of the expectations survey.}
  \footnotesize
  \centering
  \label{tab:demography}
  \begin{tabular}{l|r|r|r}
    \hline
    & \# respondents & Gender & Age (Years)\\
    &               &             & 10--29~/~30--49~/~50--\\
    \hline\hline
    Facebook  &  198 & F:54 M:46 (\%)&  31~/~60~/~9 (\%)\\
    \hline
    Twitter   &  170 & F:56 M:44 (\%) &  41~/~51~/~8 (\%)\\
    \hline
  \end{tabular}
\end{table}

Next, to answer the ``why'' question we recruited participants to take an online survey. 
As summarized in Table~\ref{tab:demography}, the demography of the respondents shows that responses represent a diverse, cross-section of respondents. 
Key findings derived from the closed-ended questions are as follows: 
(1) 52.3\%/41.4\% of Twitter/Facebook users responded that they have used the blocking mechanism; 
(2) 92.4\%/93.9\% of Twitter/Facebook users responded that social web service should {\em not} limit the number of accounts a person can block on the service. This result indicates that users do not expect to have limitations on the number of blockable users. 
We also included the open-ended questions: ``why do you block other users?'' and ``why do you think that there should be {\em no} limitation on the number of blocked users?''
Typical answers to the first question include ``do not want to read the unwanted messages/posts'' and ``not to be tracked by strangers/trolls/ex-friends/coworkers, etc.''
Typical answers to the second question include ``there are a huge number of spam/bogus accounts'' and ``just adding unwanted users to the blocklist is easy to maintain.''

The observations derived from the web service log analysis and the online survey imply that simply disabling our side channel, user-blocking, is {\em not} a desirable countermeasure against the threat from the viewpoints of actual usage of a service and users' expectations.

\section{Attack Overview}
\label{sec:threat}

In this section, we give a brief overview of the attack. We present the threat model and the attack flow with a concrete example. We also elaborate on the novelty of the attack and how it compares to some of the existing works in this area.

\subsection{Threat Model}
\label{sec:threat-model}
In this attack, the attacker's goal is to determine the social account of the visitors to her/his website. We present two possible attack scenarios under this goal. In the first, the attacker targets unspecified mass users in order to determine who visited the attacker's website, for the purpose of, e.g., marketing. In the second scenario the attacker targets a limited number of users with already known identities, such as their names or email addresses, and wants to determine their anonymous accounts which cannot be searched for using such identities. In both scenarios, the visitor's privacy is obviously breached, as the identity of the user or their private activity is revealed to the attacker without their consent. 

Our attack employs a {\em cross-site timing attack}, which is an attack that combines
cross-site request forgery (CSRF) and a timing attack~\cite{Bortz_WWW2007}. 
Cross-site timing attacks bypass the same-origin policy and
enable an attacker to obtain information using the target's view of
another site, i.e., in our context, the attacker can know whether or
not the target user is blocked by the attacker-prepared {\em signaling accounts}.
As we detail in Section~\ref{sec:rtt}, the status of
blocked/non-blocked can be estimated from the time a web server of a social web
takes to load a web content, or the round-trip time (RTT), of the profile page of a signaling account.
As such, we make the following assumptions, which we will discuss in additional detail in Section~\ref{sec:discussion}.

\smallskip\noindent\textbf{Attack Trigger.} We assume that the attacker can somehow induce their target to visit a malicious
website.
For example, the attacker uses malvertising techniques~\cite{Li_CCS2012} or
simply sends out email messages, in which case they can also link e-mail addresses to social accounts.
Further details on this are discussed in Section~\ref{sec:discussion}.\\
\smallskip\noindent\textbf{Log-in Status.} We assume that a target person has logged into the social
web services, i.e., that cookies are enabled on the person's web browser.
This assumption plays a vital role in the success of the attack
because the logged-in status triggers the difference between views of
profiles of blocking and non-blocking accounts.
Because the majority of web services, e.g., Facebook, have an automatic sign-in option,
we consider this assumption to be reasonable.\\
\smallskip\noindent\textbf{User Device.} We assume that the target person uses a PC when accessing
the malicious website.
This premise covers more than 70\% of social web service users~\cite{GLOBALWEBINDEX}.
Users of mobile platforms typically access social
web services through dedicated mobile apps instead of the
web interface provided for mobile browsers. 
Therefore, we cannot easily apply the attack to a mobile device.

\subsection{Attack Flow and Example}

\begin{figure*}[tbp]
  \begin{center}
    I. Side-Channel Control Phase \\
    \includegraphics[width=120mm,clip]{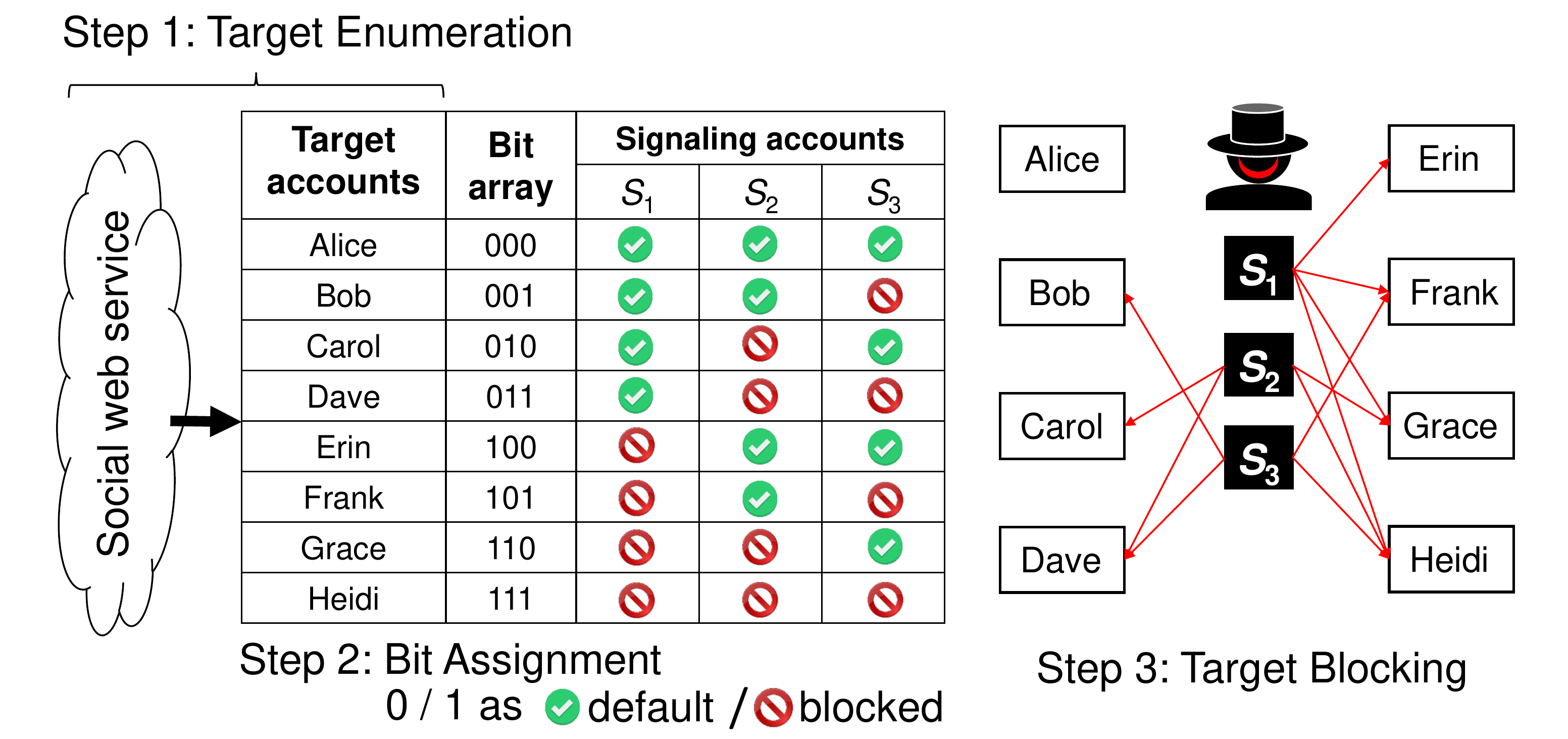}
    \vspace{5mm} \\
    II. Side-Channel Retrieval Phase \\
    \includegraphics[width=120mm,clip]{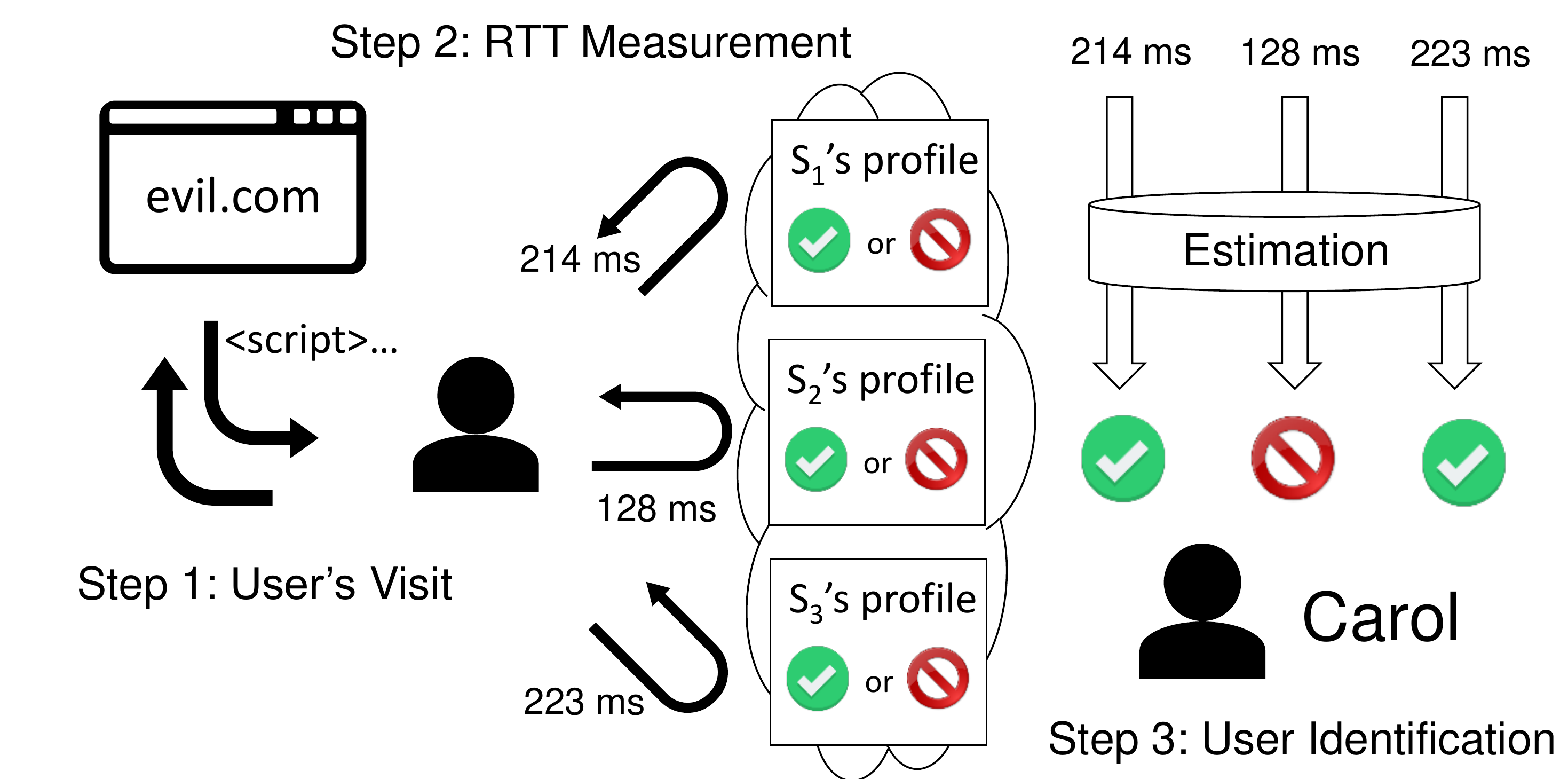}
    \caption{Attack overview}
    \label{fig:blocking_table}
  \end{center}
\end{figure*}

As illustrated in Figure~\ref{fig:blocking_table}, our attack has two separate phases: the side-channel control phase and the side-channel retrieval phase. 
Below, we describe the steps in each phase with a concrete example. Note that some details are omitted for simplicity but will be described in later sections.

\vspace{2mm}
\par\smallskip\noindent\textbf{I. Side-Channel Control Phase} 
\vspace{2mm}

The purpose of the side-channel control phase is to construct user-identifiable side-channel data through user blocking. This phase is required just once before performing the attack.

\par\smallskip\noindent\textbf{Step 1. Target Enumeration: } 
For a social web service of interest, the attacker first enumerates the users who will be the target of the attack. Let $N$ be the number of targets. The attacker can target either mass (randomly sampled or even all) user accounts or a limited set of selected users (e.g., celebrities, high-level corporate officers) according to the attacker's purpose. Large-scale, efficient enumeration of social accounts can be achieved in several  ways, as in~\cite{Krishnamurthy_WOSN2008, Dey_Infocom2012}. Note that because this attack leverages CSRF, whether the account is closed (e.g., private, protected) is not a concern as long as the account is blockable.

\noindent\textit{In the example}, the attacker lists a small set of $N = 8$ users who will be the target of the attack. If the attack succeeds, the attacker will be able to identify the accounts of these eight users whenever they visit the attacker's website while logged onto the social web services.

\par\smallskip\noindent\textbf{Step 2. Bit Assignment: }
The attacker prepares $m$ accounts on the social web service where $m$ is a number satisfying $2^m \geq N$; these accounts are referred to as ``signaling accounts'' and denoted as $S_i$, $i=1\ldots m$. The attacker encodes a set of target users into bit arrays with length $m$, with the value of the $i$-th bit of each array corresponding to ``block'' ($1$) or ``do not block'' ($0$) by account $S_i$. The attacker can express a maximum of $2^m$ distinct target users, but at the cost of increase in $m$, the attacker can further add redundant bits to produce an error-correcting code. 

\noindent\textit{In the example}, the attacker prepares $m = 3$ signaling accounts, $S_1$, $S_2$, and $S_3$, with each target user is mapped into distinct bit arrays of length $m$, as shown in the table. All possible bit patterns are mapped to the users and there are no redundant bits.

\par\smallskip\noindent\textbf{Step 3. Target Blocking: }
The attacker controls the signaling accounts to block each target user according to the bit array. Note that the number of blocking that must be performed per signaling account is approximately half of the total number of targets, as shown in the figure. It is not difficult to see that this requirement can be controlled at the cost of adding more redundant signaling accounts, i.e., the block/non-block table in the figure will become more sparse.

\noindent\textit{In the example}, $S_1$ is configured to block Erin, Frank, Grace, and Heidi, with the remaining four users left non-blocked (default). $S_2$ and $S_3$ are configured in a similar manner.

\vspace{2mm}
\par\smallskip\noindent\textbf{II. Side-Channel Retrieval Phase} 
\vspace{2mm}

The purpose of the side-channel retrieval phase is to identify the user utilizing the data retrieved through the timing side channel. This phase is executed every time a user accesses the attacker's website.

\par\smallskip\noindent\textbf{Step 1. User's Visit: }
When a user visits the web server under the control of the attacker, JavaScript code is downloaded and is executed on the user's browser. 

\par\smallskip\noindent\textbf{Step 2. RTT Measurement: }
The JavaScript code (as detailed in Bortz~\cite{Bortz_WWW2007}) measures the time taken to load the profile of the signaling accounts by sending HTTP requests to each of these accounts. 
Note that, as this is a CSRF, the request is issued on behalf of the user's account. Special RTT measurements are also performed to determine the threshold value used in the next step, but we omit the details here.

\noindent\textit{In the example}, the script issues HTTP requests to the profile page of each of the signaling accounts --- $S_1$, $S_2$, and $S_3$ --- and receives the measurements of $214$, $128$, and $223$ \rm{ms}, respectively. 

\par\smallskip\noindent\textbf{Step 3. User Identification: }
The attacker then tries to identify the user from the measurements acquired in the preceding step. Because the time needed to load the profile of a blocking account exhibits a statistical difference from that needed to load the profile of a non-blocking account, the sequence of measured time samples can be used to build a bit array of ``blocked'' and ``non-blocked'' states. Once the bit array is recovered, the attacker does a lookup on the bit array map and identifies the user.

\noindent\textit{In the example}, the measurements, $214$, $128$, $223$ \rm{ms} are compared against a threshold value of, say, $150$ \rm{ms}, and are determined to be non-blocked, blocked, and non-blocked, respectively. This result is represented as a bit array $\{010\}$, enabling the attacker to infer from the table that the user who visited the malicious site is Carol\footnote{As we will detail in Section~\ref{sec:attack}, when $\{000\}$ is observed, it is still possible to distinguish Alice from non-target users by using two special accounts that do/don't block all the target users.}.

\subsection{Novelty of the Attack}
\label{sec:novelty}
While our attack is certainly novel overall, its conceptual novelty lies primarily in the side-channel control phase rather than in the side-channel retrieval phase, which can be implemented using many different existing approaches in addition to that adopted in our implementation~\cite{Bortz_WWW2007}. The side-channel control phase is made particularly novel by its use of the underlying concept of {\em visibility control}, which allows for the encoding and retrieving of arbitrary bits of data independent of what the side channel is. This flexibility inherently enables the attack to achieve account identification in a generic manner. By contrast, most similar methods that exploit browser side channels focus on stealing the content of a specific resource, limiting the acquirable data to that related to the targeted resource. Rather than studying such resource-specific side-channel acquisition methodologies, we questioned and exploited the design of general systems equipped with visibility-control features, e.g., user blocking. To the best of our knowledge, this concept has not been previously discussed in the literature despite its significant potential impact on nearly all major social web services currently operating.

We now compare our work to two of the major recent studies in this area. The goal of the first study was to retrieve various user data (e.g., age, contacts, search history) through several browser side-channel techniques~\cite{Goethem_CCS2015}. The major difference between this work and ours is that it was somewhat focused on the development of individual techniques to acquire resource-specific side channels. Although this makes their methodology more powerful in the sense that it can even reveal a user's private information (e.g., search history), their methodology and goals were more service- and resource-specific. By contrast, the purpose of our work is to find user accounts and then link these with {\em all} available public information to which they are tied  independent of the target resource used for sending side-channel data. Another similar study involved an attack based on browser history stealing~\cite{Wondracek_SP2010}, which, in the authors' words, shared a goal similar to ours of user identification or de-anonymization. This approach exploited the (now eliminated) mechanism allowing an attacker to infer a user's browser history to determine if the user belongs to certain groups based on the presence of access history to certain pages. Methodology-wise, the concept of repetitively identifying the groups to which a target user belongs, until to the point where the target can be uniquely identified, is conceptually similar to our approach. The main difference, however, is that our method allows for the construction of such groups in advance in an arbitrary manner. Thus, while our approach requires some initial setup effort, it has the advantage of being much more reliable in assuring identification (i.e., no ambiguity remains due to a lack of groups) as long as the side channel can be correctly retrieved.

\section{User-blocking Side Channel}
\label{sec:rtt}

This section aims to demonstrate that the differences between the
time to load profile pages of blocked and non-blocked users can
be used to perform a timing attack.
In the following, we first look at the characteristics of the RTTs
measured for blocked and non-blocked accounts.
Next, we present several techniques that can increase the
distinguishability of RTTs.
Finally, after applying the RTT expansion techniques, we test whether
the RTTs are statistically distinguishable using various social web
services, which include popular social media such as Twitter and
Facebook and other web services such as eBay and XBox Live. 

\subsection{Characteristics of RTTs}

\begin{figure}[tbp]
  \begin{center}
  {\footnotesize
  (a) Facebook \\
  \includegraphics[width=120mm,clip]{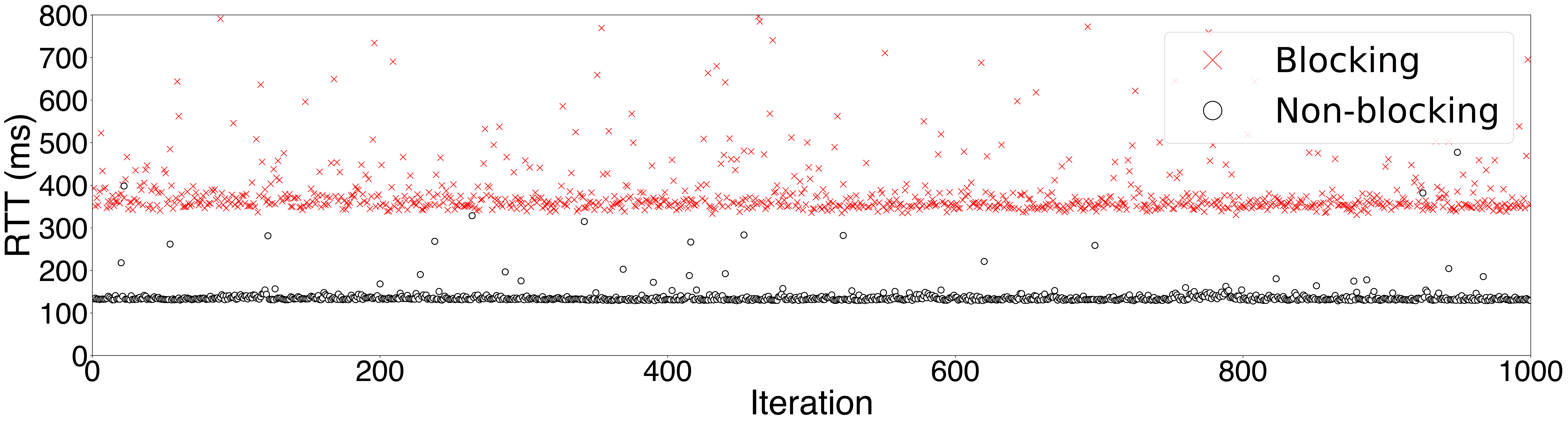} \\
  (b) Twitter \\
  \includegraphics[width=120mm,clip]{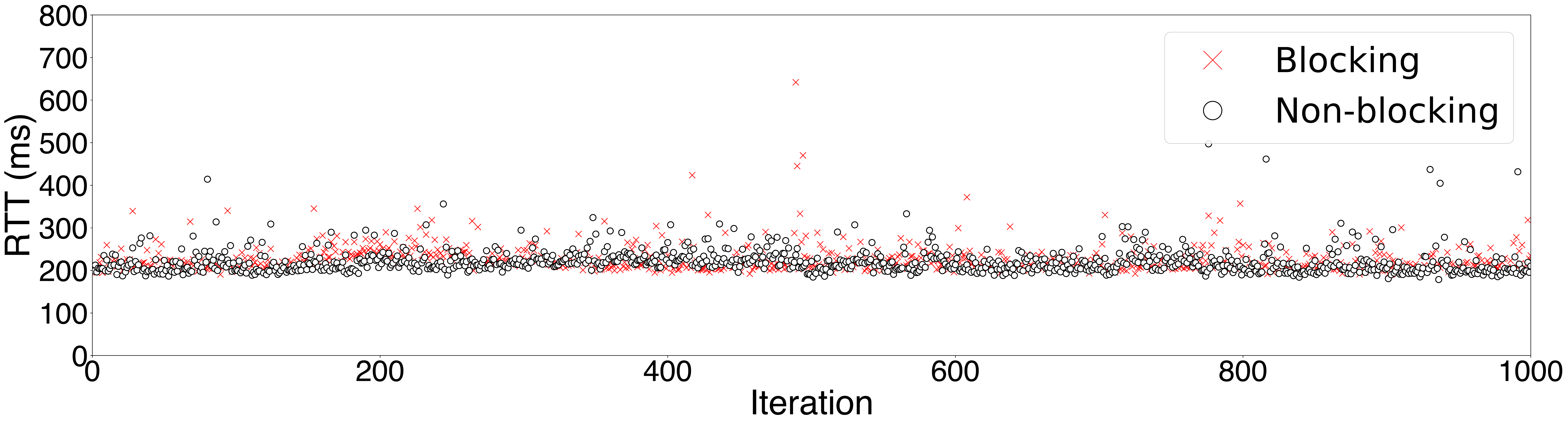}\\
  (c) Tumblr \\
  \includegraphics[width=120mm,clip]{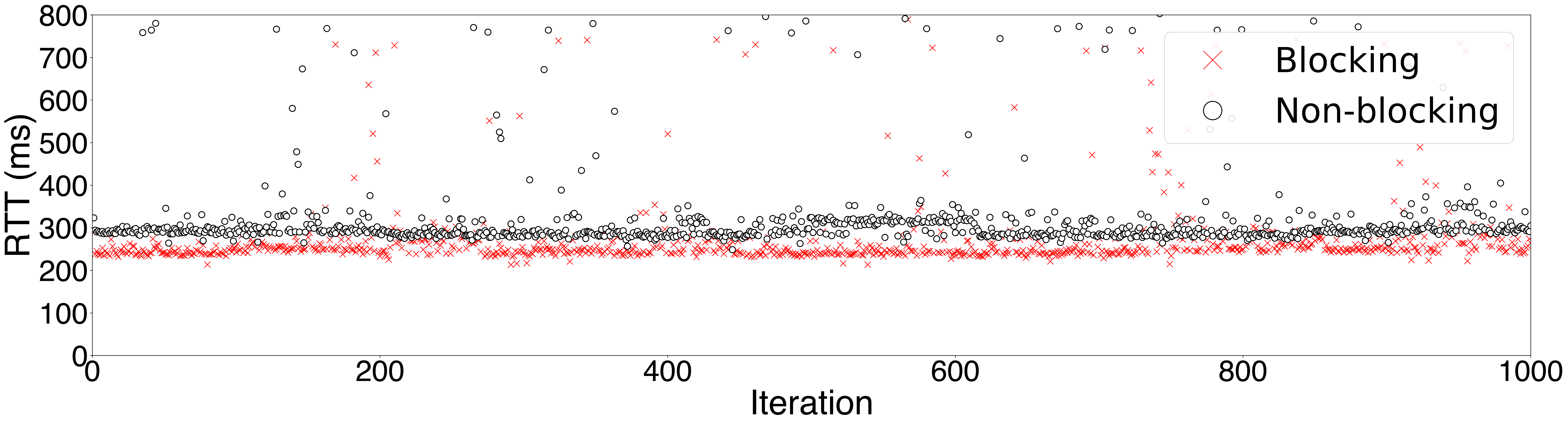}\\
  \caption{{\footnotesize Distributions of RTTs for blocking and non-blocking accounts.}}
  \vspace{-2.1mm}
  \label{fig:dist-of-rtt}
  }
  \end{center}
\end{figure}

Here, we briefly describe the setup for our experiments. We executed a simple JavaScript code on a browser logged-in to a service with account A. The JavaScript issues GET requests to a page associated with an account which blocks A, and another page associated with an account that does not block A. Our objective is to see whether we can see the differences in the RTT measurements associated with these two types of accounts: blocking and non-blocking.

In the following, we characterize the measured RTTs using three social
web services, Facebook, Twitter, and Tumblr as the representative
examples.
We study other services in the next subsection. 
Figure~\ref{fig:dist-of-rtt} shows the distributions of the
measured RTTs.
For Facebook, there is a clear gap between
the RTT distributions for blocking and non-blocking accounts. 
For Tumblr, even though two distributions are closer,
we see the difference between the distributions.
We study whether or not this slight differences can be used as
the timing side channel in Section~\ref{sec:experiments}.
For Twitter, the distributions suggest that there is no sufficient difference to distinguish their RTT difference. Nevertheless, we have discovered that it is possible to intentionally amplify their RTT difference by posting more content to the profile page. More details on this will be described in the next subsection.

Note that, while we see longer RTTs for non-blocking accounts
on Tumblr, we see longer RTTs for blocking accounts on
Facebook.
It is natural that the profile pages of blocking accounts are loaded
quickly because the content of these pages may be lighter than those of
the profile pages of non-blocking accounts. 
While not conclusive, we conjecture that this could be because Facebook does not utilize its server-side on-memory cache at all when generating content for the case of blocked.
In either case, we can distinguish between the blocked and
non-blocked states using the RTT measurements.

\subsection{Improving RTT Distinguishability}
\label{sec:teqs_dist}
We present three techniques that can make the differences in the RTTs
more prominent, i.e., these are the ways to make the timing attack
more successful.

\begin{figure}[tbp]
  \begin{center}
  {\footnotesize
  Twitter (profile page content filled) \\
  \includegraphics[width=120mm,clip]{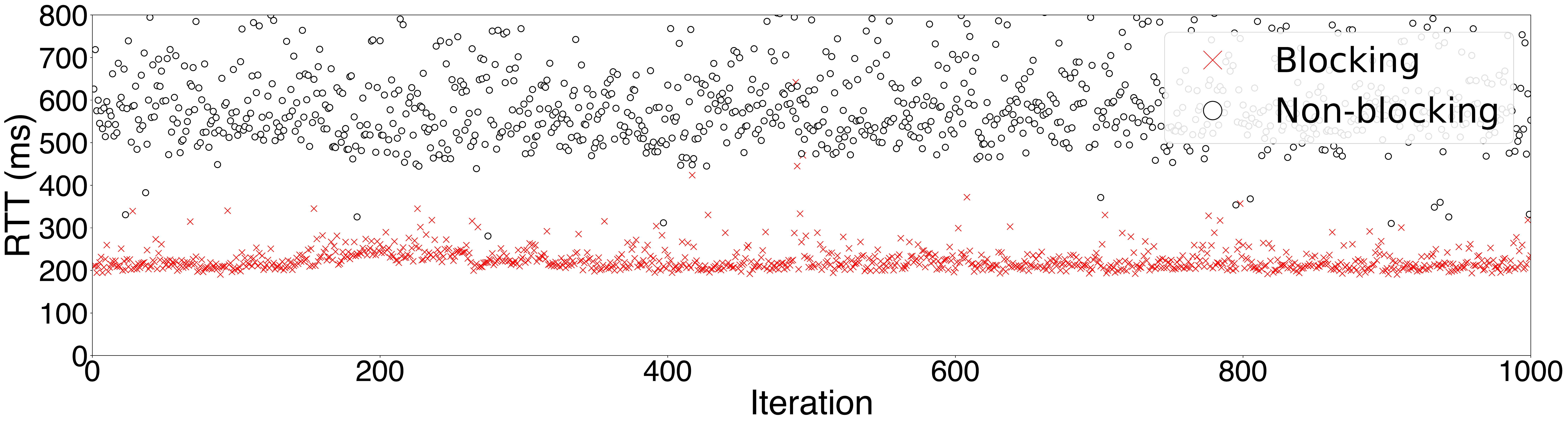} \\
  \caption{{\footnotesize Distributions of RTTs for blocking and non-blocking accounts, after filling the Twitter user profile page with content}}
  \label{fig:dist-of-rtt_filled}
  }
  \end{center}
\end{figure}

\smallskip\noindent\textbf{Change of content size.}
The first technique is to place as much information as possible on
the user profile pages of the signaling accounts. 
This technique can increase the time to load the profile page when
the signaling account of the page is visible to the target, i.e., the
signaling account does not block the target. 
We performed a simple experiment using Twitter. We prepared two Twitter
accounts, one with the default setting and another with the
maximum amount of content (texts and URL links) that appears on the
profile page. 
Figure~\ref{fig:dist-of-rtt_filled} shows the RTT distributions after filling the profile page with large amounts of content. Comparing this with Figure~\ref{fig:dist-of-rtt} (b) which shows the RTT distributions before adding the content, we now have a clear difference between blocked and non-blocked RTTs, suggesting that this technique can dramatically improve their distinguishability.

\smallskip\noindent\textbf{Use of different pages.}
Another technique is to make use of various pages other than the user profile page. In many cases, the page subject to blocking is the user profile page, which displays the user's basic information or recent posts. However, depending on the service or their implementation, there are cases where observable differences do not appear on the profile page but do appear on other pages. For example, on eBay, a user cannot prohibit another user from accessing their profile page; however, a user can prohibit another user from bidding on the items they list. In other words, the content on the item page would yield a difference depending on whether the viewing user is blocked by the owner of the item. Leveraging this fact, by preparing an item beforehand and making the victim send requests to the item page instead of the profile page, the attacker would be able to observe the RTT difference required for the attack. 

Similarly, Flickr does not prohibit a blocked user from viewing the blocker's profile page, but it does prohibit the blocked user from sending a message to the blocker. More specifically, there is a page for sending a messages to other users and, if the sender is not blocked from the receiver, a text area and a submit button are displayed on the page; however, if blocked, these objects are not shown and a warning message is displayed. Such a difference may also yield the RTT difference necessary for our attack.

In addition, some pages with AJAX-based implementation have a structure where after requesting and rendering the initial HTML content, they request additional content, e.g., a JSON content, from another URL using JavaScript's XMLHttpRequest. In some services, the blocked/non-blocked difference is only present in the JSON data that is acquired afterwards, instead of in the HTML content acquired first. The problem with this situation is that the RTT measurement script used for cross-site timing attacks does not actually render the acquired page content; therefore, the RTT of the content acquired afterward from JavaScript cannot be measured. In such cases, the attacker must directly send requests to the URL for the JSON data. In our investigation, we found that Tumblr and Xbox.com had this structure, but we were able to make the attack feasible by switching the request destination to the JSON URL instead of the HTML URL.

\subsection{Distinguishability of RTTs}

We tested whether the RTTs for blocking and non-blocking accounts were
statistically distinguishable.
To this end, we leveraged the Mann-Whitney U test, which is a
nonparametric statistical test used to compare differences between two
independent samples; it tests whether a randomly selected value from
one sample is less than or greater than a randomly selected value from
another sample. 
For our experiments, we picked 16 popular social web services. 
For each service, we measured the RTTs between blocking/non-blocking
accounts and the blocked account.
We applied the Mann-Whitney U test and computed the $p$-values.
The results are summarized in Table~\ref{tab:service_summary}.
The results show that all services have low $p$-values and imply that
the distributions are distinguishable in $12$ out of $16$ services when the
significance level is $0.01$.

\begin{table}[tbp]
  \caption{Social web services with user-blocking mechanism. \footnotesize{$\Delta_{0.05}$ shows the difference in $5$-percent tile values for blocked/non-blocked RTT measurements. Dist. is the distinguishablity showing Y when the $p$-value less than $0.01$. \# of users are from various web resources as of May 2017}}
\tiny
\centering
  \label{tab:service_summary}
  \footnotesize
\scalebox{0.95}[0.95]{
\begin{tabular}{l|l|r|r|r|c}
    \hline
    Service & Category & \# users & $\Delta_{0.05}$ & $p$-value & Dist.\\

    \hline\hline
    Facebook& Social & 1.96B & 212 ms & $<$0.0001 & Y \\
    Instagram & Photo & 700M & 29 ms & $<$0.0001 & Y \\
    Tumblr & Microblog & 550M & 43 ms & $<$0.0001 & Y \\
    Google+ & Social & 540M & 1,080 ms & $<$0.0001 & Y\\
    Twitter & Microblog & 328M & 312 ms & $<$0.0001 & Y \\
    eBay& Shopping & 167M  & 589 ms & $<$0.0001 & Y\\
    PornHub & Porn & 75M  & 9 ms & 0.0034 & Y \\
    Medium & Forum & 60M & 332 ms & $<$0.0001 & Y \\
    Xbox Live & Game & 52M  & 110 ms & $<$0.0001 & Y \\
    Ashley Madison & Dating & 52M & 8 ms & 0.0097 & Y \\
    Roblox& Game & 48M & 98 ms & $<$0.0001 & Y \\
    Xvideos& Porn & 47M & 16 ms & $<$0.0001 & Y \\
    Quora& Forum & 190M & 5 ms & 0.4561 & N \\
    Flickr& Photo & 122M  & 1 ms & 0.2678 & N \\
    DeviantArt & Art & 65M& 11 ms & 0.0674 & N \\
    Meetup& Social & 30M  & 9 ms & 0.3878 & N \\
    \hline
  \end{tabular}
}
\end{table}
\section{User Identification Attack}
\label{sec:attack}

In this section, we first formulate the user identification attack,
which works on the basis of the two building blocks, user-blocking and
cross-site timing attack.
The attack introduces two functions, encoding and
decoding, which are the functions an attacker can arbitrarily set to
map target users and leaking information (RTTs).
Next, we describe the techniques we developed for the timing attack. 
Finally, we present two extensions of the attack. These extensions aim to
make the attack more successful. 

\subsection{Formulation}

Let $m$ and $N$ denote the numbers of the signaling and target
accounts, respectively. 
We configure $m$ as the
minimum integer value that satisfies $2^m \geq N$. 
If an attacker wants to target one million of accounts, $m$ is
configured to $m=20$.

In the setup phase, an attacker creates a table that maps target user
accounts to bit arrays with a length of $m$. 
Let $U_i\quad(i=1,\ldots,N)$ be the target user accounts.
For each $U_i$, the table has a bit array entry, $B_i=\{b_1 b_2\ldots
b_m\}$, where $b_j\in \{0,1\}$ corresponds to a bit.
We refer to the rule that maps $U_i$ into $B_i$ as {\em encoding}, i.e.,  
\begin{equation*}
  B_i = \mbox{encode}(U_i).
\end{equation*}
Next, we configure the signaling accounts, $S_j\quad(j=1,\ldots,m)$ as
follows.
Let $\theta_{ij}\in\{0,1\}\quad(i=1,\ldots,N, j=1,\ldots,m)$ be an indicator
function that satisfies 
\begin{equation*}
  \theta_{ij} = 
  \begin{cases}
    1 \quad\mbox{if $b_{ij}=1$ },\\
    0 \quad\mbox{else},
  \end{cases}
\end{equation*}
where $b_{ij}$ is the $j$-th bit of the bit array $B_i$.
Then, for each signaling account, $S_j$, the account is configured to
block the user $U_i$ if $\theta_{ij}=1$.
Because each bit takes the value $b_{ij}=1$ with a probability of $0.5$,
each signaling account needs to block approximately $N/2$ target
accounts. 
One may instantly come up with a defense that poses a limit on the number of user-blocks an account can have. To thwart such a countermeasure, we propose a technique described in Section~\ref{sec:attack_extension}.

In the attack phase, the attacker sets up a malicious website and lets
target users access it, following our threat model. As described
in the previous section, using the timing attack, the website can
secretly measure RTTs for the $m$ of signaling accounts. Note that
measurements can be parallelized to speed up the process.
Let $\mathbf{R}_{j}=\{R_1,R_2,\ldots\}$ be the sequence of RTT
measurements obtained for the signaling account $S_j$. 
Using the techniques that will be described in the next subsection, we
estimate whether or not the target user is blocked by $S_j$. 
Let $\widehat{b_j}  \in \{0,1\}$ denote the estimate of
the blocked/non-blocked ($1/0$) from the RTT measurements, i.e.,
\begin{equation*}
  \widehat{b_j} = \mbox{est}(\mathbf{R}_j).
\end{equation*}
Using the entire estimates, we have the estimate of $B$ as 
$\widehat{B}=\{\widehat{b_1}\ldots\widehat{b_m}\}$.
Finally, we identify the target user using the table created in the
setup phase; i.e.,
\begin{equation*}
  \widehat{U} = \mbox{decode}(\widehat{B}). 
\end{equation*}

In the next subsection, the estimation, 
$\widehat{b_j} = \mbox{est}(\mathbf{R}_j)$, is described in detail.

\subsection{Estimating Blocked/Non-blocked Status}
\label{sec:detecting-block}

Prior to the actual attack, we determine whether or not a visitor of
the website has been included in the target list, i.e., we employ a
membership test.
To this end, we prepare the following two reference accounts: a
{\em closed account}, which blocks all users included in the list of
target users, and an {\em open account}, which does not block any users
at all.
We first measure the RTTs for each of the closed and open accounts.
The measurements consist of $k_0$ trials for each account, 
where we use multiple trials because the decision based on a one-shot
measurement may have errors due to jitter in the RTTs.
The idea is to compare the measured RTTs for closed/open
accounts to see if they are significantly different.
If we observe a significant difference, we can conclude that the
visitor has been listed and continue the attack; otherwise, the
visitor has not been listed and the attack procedure is terminated.

To determine if the measured RTTs are for the closed or
open accounts, we again leverage the Mann-Whitney U test.
Because the computation of the U test is simple and lightweight, the
membership test can be completed immediately after we collect the RTTs.
In this study, we adopted a significance level of $\alpha = 0.01$.
We also need to configure the parameter $k_0$. 
As shown in the next section, we empirically derived a
conservative value of $k_0$ as $k_0=30$, which worked for various social
web services. 

After the attacker determines that the visitor is likely listed, the
attacker moves to the next step.
Let $C_{0.05}$ and $O_{0.05}$ be the 5th-percentiles of the RTT values
measured for the closed and open accounts, respectively.
We adopt the 5th-percentile as the threshold to eliminate
outliers. Note that, even though we could use the minimum values for the RTTs
as does the pathchar algorithm does~\cite{Jacobson_Pathchar}, we observed that the
RTTs could include small outliers, which could be caused by
server-side mechanisms such as data caching or load balancing.
These values are used as the thresholds to estimate the blocked /
non-blocked state, i.e., for a measured RTT sequence for a signaling
account $S_j$, we compute the 5th-percentile of $\mathbf{R}_j$ as
$R_{0.05j}$.
We do not necessarily make $k$, the number of trials $S_j$, equal to $k_0$.
An attacker can adjust the $k$ according to the his/her requirements for
the trade-offs between accuracy and speed.
If the obtained  $R_{0.05j}$ is closer to $C_{0.05}$, the attacker
estimates the visitor has been blocked by the signaling account $S_j$.
Otherwise, s/he estimates the visitor has not been blocked by the
signaling account; i.e., 
\begin{equation*}
  \widehat{b_j} = 
  \begin{cases}
    1 \quad\mbox{if}\quad |R_{0.05j} - C_{0.05}| < | R_{0.05j} - O_{0.05}|,\\
    0 \quad\mbox{else}.
  \end{cases}
\end{equation*}

By continuing this process for all $j\in\{1,\ldots,m\}$, the attacker
can estimate the bit array of the visitor as
$\widehat{B}=\{\widehat{b_1}\ldots\widehat{b_m}\}$.
Finally, the bit array can be decoded into a user ID,
$\widehat{U}=\mbox{decode}(\widehat{B})$, using the procedure we have
shown in the previous subsection.
Despite the simplicity of the procedure shown above, as we show later, it can estimate the closed/open states very accurately.

\subsection{Extensions}
\label{sec:attack_extension}

Here, we introduce two extensions of the attack, {\em error-correction
coding} and {\em user-space partitioning}, which aim to further improve
the accuracy in noisy environments and to enhance the size of the target
when the number of blocks per account is limited, respectively.

\smallskip\noindent\textbf{Error-correction Coding.}
Under a stable environment, accurately classifying a bit is not difficult since sufficient significant difference between blocked/non-blocked is present. This will also be shown later in Section~\ref{sec:experiments}. On the other hand, abnormal RTTs due to some irregular factors such as temporary server overload may lead to a bit-error. Needless to say, the infrastructures used in services such as those listed in Table~\ref{tab:service_summary} which host $30$ million to $2$ billion users tend to be quite resilient against such failures; nevertheless, we can still apply error-correction algorithm in order to eliminate even the slight possibility of identification failure due to noise.

In this paper, we adopt a Reed-Solomon code, which has a high
error-correction capability and is relatively easy to implement.
In fact, as we will demonstrate later, the use of the Reed-Solomon
algorithm actually contributes to improving the estimation accuracy
in a noisy environment. 
Note that other error-correction algorithms could be used for this
purpose. To select the most suitable error-correction algorithm, one
must take into account several factors such as the error probability
distribution, the error characteristics such as bursts, and the requirements of
the available computing resources. In this paper, we are focused on the proof of
concept; therefore, we consider choosing the best error-correction algorithm to
be out of the scope of this study.

The Reed-Solomon algorithm can correct up to $K/2$ symbol errors, where
$K$ is the number of redundant symbols and $r$ (bits) is the size of
the symbol.
Because the number of bits initially allocated to each user is $m$, the
number of signaling accounts that needs to be prepared by the attacker
is $m + rK$, i.e.,
the attacker needs to prepare an additional $rK$ extra signaling
accounts.
In the setup phase, the attacker first encodes the bit arrays
allocated to each user using a Reed-Solomon encoder, and then blocks the
accounts from the signaling accounts according to the bit values of
the newly encoded bit array.
In the attack phase, by decoding the bit arrays obtained via the
cross-site timing attack using the Reed-Solomon decoder, the attacker can
obtain an error-corrected bit array.

\smallskip\noindent\textbf{User-space Partitioning.}

As described in Section~\ref{sec:userblocking}, simply enforcing a limit on the number of blocks would violate a user's right to block and may result in a serious degradation of the service quality. For services that still enforce a limit despite this negative impact, the technique shown below would be effective.
Letting this limit to be $L$, the number of candidate target
users covered for identification is also limited to $L$ when using the
procedures we have introduced up to this point. To lift this
limitation, we can employ a technique we call {\em user-space
  partitioning}, which in this case splits candidate users into $S$
user spaces each containing $L$ users, allowing us to cover up to $LS$
users in total.

In the setup phase, for each user space $j\in\{1,\ldots,S\}$, an
attacker prepares a reference account that blocks all users
belonging to the $j$-th user space and the $\lceil\log_2 L\rceil$ of
signaling accounts that are used to map the targets in the space.
We also prepare the two reference accounts,
the closed and open accounts, which are used as the
basis of the RTT-based blocking/non-blocking estimation. 
In total, the number of signaling/reference accounts required is
$S\lceil\log_2 L\rceil$ and $S + 1$, respectively. 

In the attack phase, the attacker (1) identifies which user space the
target user belongs to and then (2) identifies the target in the user
space. In step (1), as in the procedures described in the previous
subsection, for each of reference account, $k$ requests are launched to
determine the user space to which the target belongs.
Note that the RTT values obtained here can be reused as the training
data in step (2). In step (2), for each of the $L$ users in the user
space found in step (1), the same identification process is performed
as explained earlier.
Note that, because we use a different set of signaling accounts for
each user space, the request URL for the cross-site timing attack must
be changed depending on the outcome of step (1); however, this can be handled
with conditional branches in the JavaScript code.

\section{Field Experiments}
\label{sec:experiments}

In this section, we perform the field experiments. 
We first evaluate the key success factor of the attack -- RTT measurement, which plays a vital role in classifying blocked/non-blocked status using the cross-site timing attack (Section~\ref{sec:accuracy}). 
Next, we evaluate the feasibility of our user identification attack; namely, we study the identification success rate (Section~\ref{sec:success-rate}) and time to complete the attack (Section~\ref{sec:cost}).

\subsection{Accuracy of Bit Array Estimation}
\label{sec:accuracy}

Due to space and time constraints, we evaluated the accuracy using
RTT values experimentally measured for the following three services:
Facebook, Twitter, and Tumblr. As shown in Table~\ref{tab:service_summary}, these
services have the top number of users and, at the same time, had no
limitations such as the limit on the number of blockable users at the
time of the experiment. In addition, as mentioned in Section~\ref{sec:rtt}, each of
these three services had different characteristics in the
blocked/non-blocked RTT difference: relatively large, medium, and
small, respectively.

The experiment was conducted by executing the 
JavaScript code on Google Chrome installed on a consumer laptop PC and
measuring the RTT.
We prepared the following three different network environments: wired
LAN, Wi-Fi, and tethering. The wired LAN and Wi-Fi are connected to a
commercial Internet Service Provider, and we assume that this is the environment of PC users who
are the main targets of our attack. Moreover, to prove that our attack
is feasible even in crude environmental conditions, we also tested the attack
on a tethering network hosted on an Android device connected to a 4G
network provided by a mobile network carrier.

\label{sec:acc_of_array}

\smallskip\noindent\textbf{Membership Test.}
We first tested the accuracy of the membership test.
We measured the RTT for each of the closed and open
accounts. As mentioned earlier, the measured RTT values are used
for (1) the membership test and (2) deriving the thresholds for the bit
classification, which will be described later.
Note that an attacker needs to calibrate the thresholds before
launching the attack because the RTT values depend on the geographical
location and network environment.

We repeated the following experiment 100 times. While logged on to a
target and non-target account, we launched $k_0$ trials for
each account and decided whether or not the account was included on the list
by applying the Mann-Whitney U test. We refer to the true
positive rate (TPR) as the ratio of correctly deciding that the target
was included in the target, and the true negative rate (TNR) as the
ratio of correctly deciding that the target was {\em not} included on
the target list.

\begin{figure}[tbp]
  \begin{center}
    \includegraphics[width=120mm,clip]{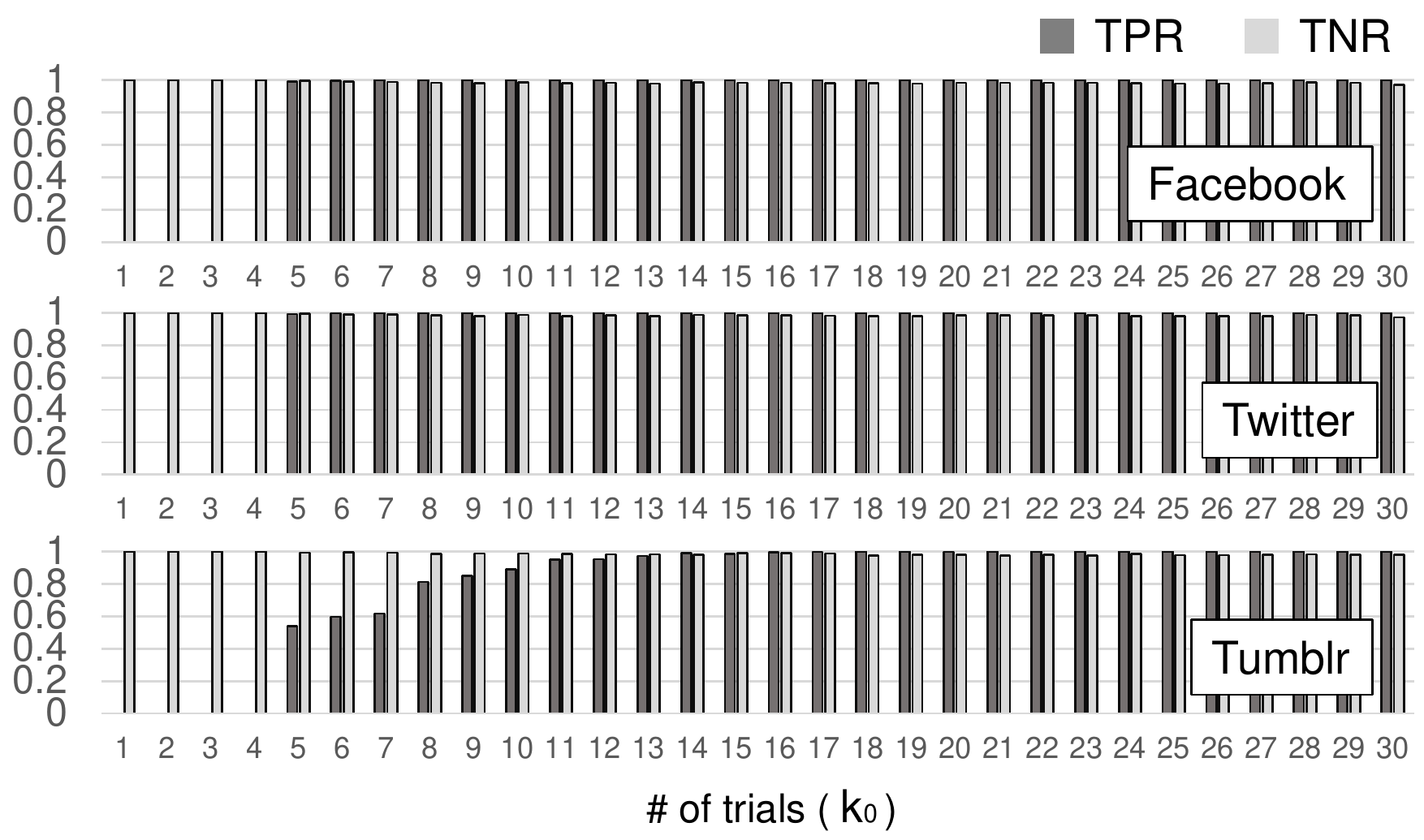}
    \caption{Relationship between the number of trials $(k_0)$, and
      TPR/TNR.}
    \vspace{-2.5mm}
    \label{fig:rtt_box}
  \end{center}
\end{figure}

Figure~\ref{fig:rtt_box} shows the relationship between $k_0$ and
TPR/TNR.
When $k_0$ is small, we have a small number of samples to estimate the
states.
Nevertheless, thanks to the strong distinguishability of the RTT
distributions, TNR was $0.97$ for all $k_0$, i.e., there were very few false
negatives, which are events where the target account was estimated
as {\em not} being listed. 
Second, for TPR, we saw degradation in the accuracy when $k_0$ was small,
especially for Tumblr. 
As $k_0$ increases, however, the TPR approaches $1.0$. 
When choosing the value of $k_0$, it is preferable that the accuracy is
consistent and that we see a sufficient difference in the samples. 
If $k_0$ is large, the accuracy will increase but the number of trials
will also increase and the time needed for an attack would become too
long. In this experiment, we empirically chose $k_0=30$, which achieved
perfect estimations for all the services.
We will use the values of $C_{0.05}$ and $O_{0.05}$ calculated from this
$k_0$ as the thresholds used in the next step.

The measured RTT values can be affected by various external
factors such as network latencies or the type of browser.
We studied how these factors affected the TPR/TNR. Table~\ref{tab:envs}
shows the results. The number of trials was set to $k_0=30$.

\begin{table}[tbp]
  \caption{TPR and TNR for under various conditions.}
  \centering
  \label{tab:envs}
  \begin{tabular}{l|r|r|r|r|r|r}
    \hline
    &\multicolumn{2}{c|}{Facebook} &\multicolumn{2}{c|}{Twitter} &\multicolumn{2}{c}{Tumblr}\\
    \hline
     & TPR & TNR & TPR & TNR & TPR & TNR\\
    \hline\hline
    Chrome/Wired & 1.00 & 0.99 & 1.00 & 0.98 & 1.00 & 0.99\\
    \hline
    Wireless & 1.00 & 0.98 & 1.00 & 0.98 & 1.00 & 0.99\\
    Tethering & 1.00 & 0.98 & 1.00 & 0.97 & 1.00 & 1.00\\
    \hline
    Firefox & 1.00 & 0.98 & 1.00 & 1.00 & 1.00 & 1.00\\
    IE & 1.00 & 0.98 & 1.00 & 0.98 & 1.00 & 1.00\\
    \hline
  \end{tabular}
\end{table}

\smallskip\noindent\textbf{Single Bit Classification.}
\label{sec:single}
Next, we evaluated the accuracy of classifying a single bit into
blocking or non-blocking.
Again, we used three social web services, Facebook, Twitter, and
Tumblr.
For each service, we performed $k$ trials of RTT measurements for each
of two signal accounts with blocked/non-blocked states. 
We continued this process for 100 times and took the mean values of
the following metrics.
We refer to the true blocking rate(TBR)/true non-blocking rate(TNBR) as the rate of correctly detecting the blocking/non-blocking user as a blocking/non-blocking user, respectively.
Table~\ref{tab:acc_listing}  shows the results.
When $k\geq 20$, the detection becomes perfect for all the three
services. Moreover, in a stable environment such as Facebook/Wired, 
the classification succeeds perfectly even with $k=3$.

\begin{table}[tbp]
  \caption{Accuracy of classifying a single bit for Wired(top), Wi-fi(middle), and Tethering(bottom)}
  \centering
  \label{tab:acc_listing}
  \begin{tabular}{l|r|r|r|r|r|r}
    \hline
    & \multicolumn{2}{c|}{Facebook} &\multicolumn{2}{c|}{Twitter} &\multicolumn{2}{c}{Tumblr}\\
    \hline
    k & TBR & TNBR & TBR & TNBR & TBR & TNBR\\
    \hline\hline
    1 & 1.00 & 0.98 & 0.99 & 0.99 & 0.67 & 0.99\\
    3 & 1.00 & 1.00 & 1.00 & 0.99 & 0.89 & 0.99\\
    5 & 1.00 & 1.00 & 1.00 & 0.97 & 0.95 & 0.98\\
    10 & 1.00 & 1.00 & 1.00 & 1.00 & 0.98 & 1.00\\
    20 & 1.00 & 1.00 & 1.00 & 1.00 & 1.00 & 1.00\\
    30 & 1.00 & 1.00 & 1.00 & 1.00 & 1.00 & 1.00\\
    \hline
    1 & 1.00 & 0.98 & 0.98 & 0.99 & 0.84 & 0.99\\
    3 & 1.00 & 1.00 & 1.00 & 0.99 & 0.98 & 1.00\\
    5 & 1.00 & 1.00 & 1.00 & 0.99 & 1.00 & 1.00\\
    10 & 1.00 & 1.00 & 1.00 & 1.00 & 1.00 & 1.00\\
    20 & 1.00 & 1.00 & 1.00 & 1.00 & 1.00 & 1.00\\
    30 & 1.00 & 1.00 & 1.00 & 1.00 & 1.00 & 1.00\\
    \hline
    1 & 1.00 & 0.97 & 0.98 & 0.99 & 0.68 & 0.99\\
    3 & 1.00 & 0.99 & 1.00 & 0.98 & 0.92 & 0.99\\
    5 & 1.00 & 0.98 & 1.00 & 0.97 & 0.98 & 1.00\\
    10 & 1.00 & 1.00 & 1.00 & 1.00 & 1.00 & 1.00\\
    20 & 1.00 & 1.00 & 1.00 & 1.00 & 1.00 & 1.00\\
    30 & 1.00 & 1.00 & 1.00 & 1.00 & 1.00 & 1.00\\
    \hline
  \end{tabular}
\end{table}
\vspace{-2.5mm}

\subsection{Attack Success Rate in the Wild}
\label{sec:success-rate}

We now show the result of our experiment conducted in an
environment imitating an actual attack scenario in the wild. We set
the length of a bit array to $m=24$, which can cover over $16$ million
users. In addition, we applied a Reed-Solomon code with a block length of $4$ bits
with eight redundant bits, which enables it to correct one
block of error. According to the above setting, we prepared
$34$ accounts in total,  which included $32$ signaling accounts, a closed account,
and an open account, with the appropriate blocking done against the users
on the target list.

Regarding the targets, we assigned a random bit array of length
$24$ to each of the $10$ social accounts we actually own. We encoded
these bit arrays using the Reed-Solomon code and calculated the bit
arrays assigned as the redundant bits. We prepared $10$ additional
accounts which are not included in the list.  For each of the $10$
accounts on the target list and the $10$ accounts on the non-target
list, we logged in to and accessed the attacker's website and
evaluated if the account was correctly identified. We repeated the
visit two times per account, resulting in a total of $40$ identification trials.

As the parameters for the number of trials, we selected $k=30$,
which we experimentally determined yielded good accuracy. The
service and network environment pairs we chose were Facebook/Wired LAN,
Twitter/Wireless LAN, and Tumblr/Tethering. We refer to 
the TPR as the rate of correctly identifying a target to
be included on the list, and the TNR as the rate
of correctly identifying a non-target to be not included on the
list. In addition, of the users who were identified as being included on the
target list, we refer to the identified rate (IDR) as the rate of
correctly identifying the user without the error-correction code, and
refer to the identified rate with error correction (IDR/EC) as a
similar figure but with error correction.
In Table~\ref{tab:wild_acc}, we show the classification accuracy we obtained in this
experiment.

\begin{table}[tbp]
  \caption{Accuracy of the User Identification Attack.}
  \centering
  \label{tab:wild_acc}
  \begin{tabular}{l|r|r|r}
    \hline
     & Facebook/wired & Twitter/WiFi & Tumblr/tethering\\
    \hline\hline
    TNR & 1.00 (20/20) & 1.00 (20/20) & 0.95 (19/20) \\
    TPR & 1.00 (20/20) & 1.00 (20/20) & 1.00 (20/20) \\
    \hline
    IDR & 0.95 (19/20) & 1.00 (20/20) & 1.00 (20/20) \\
    IDR/EC & 1.00 (20/20) & 1.00 (20/20) & 1.00 (20/20) \\
    \hline
  \end{tabular}
\end{table}

The result shows that the experiment succeeded with extremely high
accuracy. This was expected from the good results we obtained
from the experiments in Section~\ref{sec:acc_of_array}. 
For Facebook/Wired, there was one failure case which identified the target as a wrong user. Examining the network log for this case revealed that some requests to one of the signaling accounts had returned 502 response code due to temporary server error. Our script measures the RTT even if an error code is returned, but since no content is returned, the response time would not likely be the one desired. This occurred with $3$ of the requests over only $1$ second of duration, but the RTT value had dropped to about 1/5 of the true RTT which was enough to cause a bit error. Nevertheless, applying the error-correction algorithm, we were successfully able to correct this bit which resulted in the success of identification. 
Note that, because we adopt the 5th-percentile, our attack is resilient to outliers which are too late, but it is prone to those which are too early.

Another case of failure was for Tumblr/Tethering, where a non-target user was incorrectly identified as a target. This is a rare case where a significant difference of around $p<0.01$ happened to occur when comparing the two sets of $30$ non-blocked requests. This example also benefited from the error-correction algorithm; without error-correction this visitor would have been identified as another user, but with Reed-Solomon code, although the error was not correctable due to too many errors, the error was still detectable. In such a case, we can still prevent mis-identification by concluding that the membership test failed and  restarting the test.

\subsection{Time to Complete the Attack}
\label{sec:cost}

\begin{figure}[tbp]
  \begin{center}
    \includegraphics[width=120mm,clip]{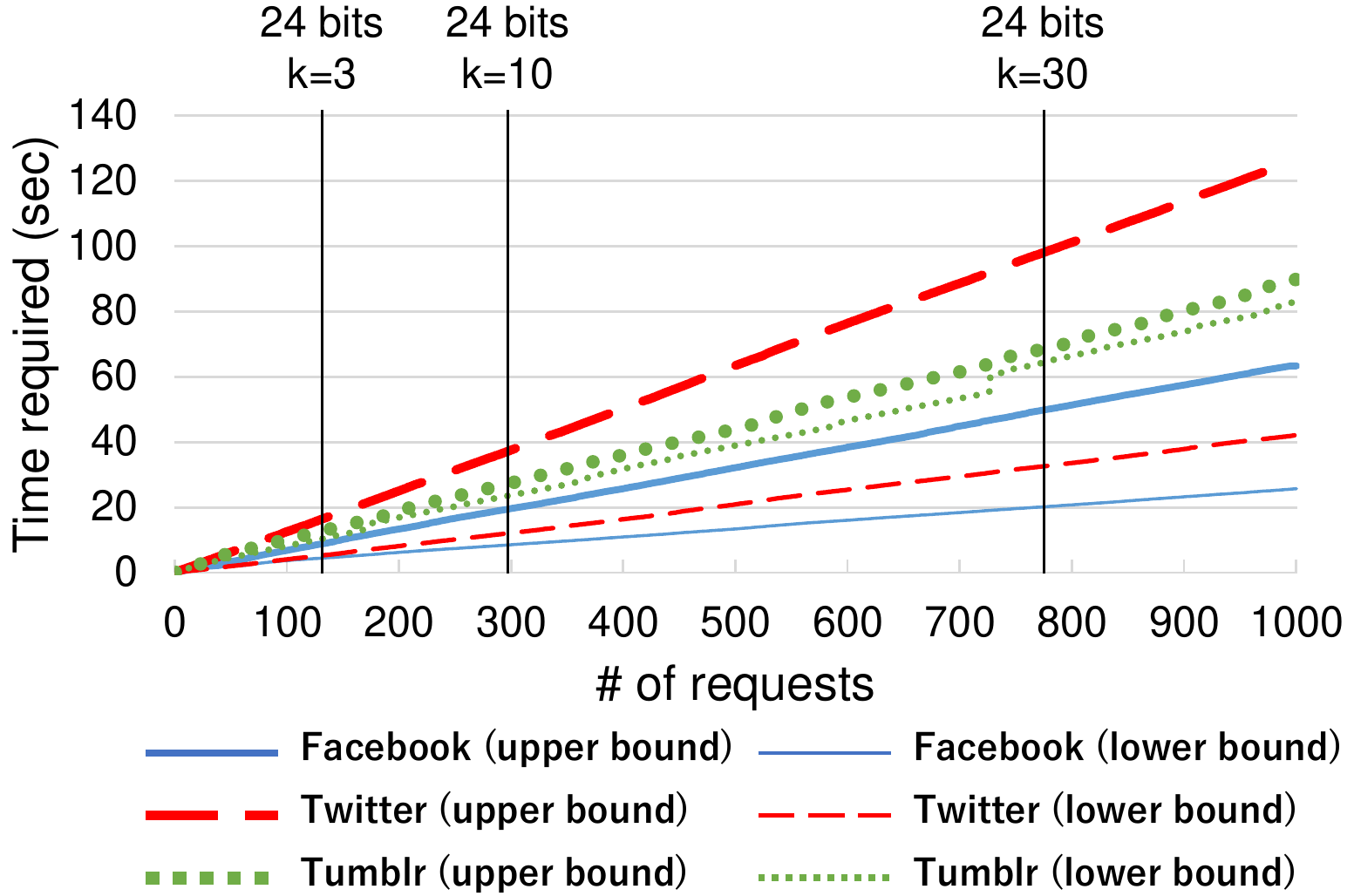}
    \caption{Number of requests vs. time.}
    \label{fig:cost}
  \end{center}
\end{figure}

The shorter the time required for the attack, the more feasible the
threat is. While the total number of requests can be calculated beforehand,
the time required to complete these trials is dependent on the
actual RTT; therefore, it needs to be evaluated experimentally. Figure~\ref{fig:cost} 
shows the relationship between the number of trials and
the required time for each service.

The ``upper bound'' value shown for each service assumes the request with
whichever has the larger of the blocked/non-blocked RTT values, that
is, it assumes the case with the longest time needed for
identification; i.e., it is the worst case. Conversely, the ``lower bound'' value assumes the
request with whichever has smaller value of the two, that is,
it assumes the case with the shortest time needed for identification; i.e., it is the most optimistic case. The number of trials issued in parallel was set to $6$, which is the
default maximum number of concurrent connections allowed on major
browsers such as Chrome, IE, and Firefox.

The total number of requests needed to make an $m$-bits decision, or
in other words, to identify the target within $2^m$ users, is 
$mk+2\times30$ when $k_0=30$. For example, for $m=24$, or targeting $16$ million users, the
total number of requests needed is $780$ when $k=30$.
This would require 20--50 seconds for Facebook, 32--98 seconds for
Twitter, and 64--68 seconds for Tumblr. According to
Table~\ref{tab:acc_listing}, in the case of Twitter, we have sufficient
accuracy even with $k=10$. The number of necessary trials is $300$ with
this setting, and the time required is 12--37 seconds.
Moreover, we can observe that we can achieve sufficient accuracy even with $k=3$ on Facebook. The total number of requests is $132$ which only takes 4--8 seconds.

\vspace{-2mm}

\section{Discussion}

\label{sec:discussion}
In this section, we discuss the attack's principle, practical aspects, known limitations, and ethical considerations.

\subsection{Principle of the Attack}

We argue that the most fundamental assumption of our attack is the presence of the {\em visibility control} property in the system, that is, 
``given a multi-user web service, there exists a way for a (rogue) user to {\em control} what other users see, individually for each user''. To be more formal, the part ``what other users see'' can be replaced with ``any observable side-effect of the system caused by a certain action taken by a user''. This assumption combined with a timing side-channel attack, which enables the attacker to steal this information from outside the system, is our attack's big picture. Because closing a side channel completely is well-known to be difficult, we believe that this visibility-control assumption is the main principle of this attack. In the case of our scenario, the ability to build signaling accounts using user blocking corresponds to this principle. 

We stress that other services under this assumption, even those without user blocking but with a similar mechanism such as group invitations or file access permissions, may also be subject to a similar class of attack. Still, the social web/user-blocking example that we used in this paper is by far the most practical application. This is likely because it satisfies several additional conditions: (1) the control can be done without the target's approval or notification and (2) the control can be done at a fine granularity, i.e., the different bits of information assignable per user is large. More specifically in our case, condition (1) is almost always achieved as an inherent nature of user blocking and condition (2) is achieved with unlimited granularity, in theory, via the creation of an arbitrary number of signaling accounts. Even though we omit further discussions concerning the presence of other such properties or the exploitability of similar systems, we believe that there is a need for further study concerning this subject.

\subsection{Practical Aspects}
Here we describe some of the applications and characteristics that extend and strengthen our attack from a practical perspective.

\smallskip\noindent\textbf{Identity Linking.}
User identification is only threatening if the identity is linked with another piece of information meaningful to the attacker. In the case of our attack, the most basic form of linking can take the form of linking the user's identity with {\em the fact that the user has visited the website} prepared for the attack. In this case, if the web content reflects the visitor's preference in any way, it may become a privacy concern. This is suitable not only for advertisement or access analyses, but also for various social engineering attacks or for blackmailing those who have accessed sites hosting pornographic content or illegal content such as pirated software. In addition, our attack can be implemented to reveal accounts on multiple services simultaneously and linking these accounts together could significantly worsen the impact of a privacy leak.

Another form of linking occurs when a person is induced to access the web server via an extra hop through another medium, resulting in a linking between the target's identity and the medium used. For example, on a social web service where the target's identity is already known, an attacker can send the target a message containing the URL of the web server. Note that this would allow the attacker to link even the web services which our attack cannot be applied to. Similarly, we can link non-web services, such as email or mobile text messages, which would result in linking an email address or phone number with a social account. Further, we can also link the target's {\em physical identity}, such as the target's physical presence or their residence, by placing or mailing a physical object, e.g., a poster or a flier, with URL, QR code or NFC tags printed on them. Note that, even though it may appear that revealing additional 
identities of a target when the target's other identities are already known is not so significant, it could lead to the identification of a target's anonymous account that cannot otherwise be discovered in a straightforward way.

\smallskip\noindent\textbf{Group Identification.}
Even though we have focused on the goal of {\em user} identification in this paper, we can easily extend this goal to {\em group} identification, that is, identifying not the user's exact identity but more general properties such as gender, nationality, or interests. The attacker could map each user to a bit array corresponding to the target attribute collected from the structured information available on the social web service. Note that this can be seen as a generalization of the user-space partitioning described previously, where a user space corresponds to a group of users with an arbitrary size mapped to a certain attribute. Group identification can be used by advertisement providers to track the visitor's attributes without unnecessarily revealing their user account. Note that the number of bits required for group identification would typically be much lower than that for user identification, making this attack significantly easier to execute than user identification.

\smallskip\noindent\textbf{Authentication-backed Identification.}
One major strength of our approach is that it is backed by the identity information guaranteed by the authentication system of the service, making it resilient against spoofing or misidentification, both of which many other methodologies suffer from. To give a simple example, when using an IP address for identification or even tracking, IP spoofing or ambiguity due to NAT or dynamic IP would interfere with this process. Note that social web service accounts are increasingly used as a building block in the modern web's authentication infrastructure. It is still possible to perform spoofing and one way is to create an account trying to mimic one's identity; however, scrutinizing the account content would usually easily reveal whether it is a spoofed account. Another way is to use a stolen account, but in this case, the victim user should be worried about much more serious problems than privacy leakage. 
In addition, because authentication is independent of the environment, it enables cross-environment (e.g., cross-device and cross-browser) identification and tracking, which is often difficult to achieve using other approaches.

\subsection{Limitations}
\smallskip\noindent\textbf{Login State Persistence.}
Our attack relies heavily on the assumption that the target user's service login state is alive while the user browses other websites. This assumption is reliant on the web cookie mechanism; therefore, the cookie's expiration time or the user configuring the browser to clear cookies on closing the browser may affect the availability of our attack. Social web services, fortunately, tend to set a relatively long or even no expiration time, as seen in the commonly available ``keep me logged in'' features~\cite{Kontaxis_Security2012}. This is likely due to the incentives to service providers from a marketing perspective, e.g., tracking and advertisement, contrary to security-critical services such as Internet banking that set a short expiration time. In addition, users would lose the convenience of being able to access the service without the need to login every time, which may be a disappointing trade-off, especially for social web services which often assumes constant usage. Note that, simply determining whether a user is logged in to certain services can be accomplished in much more lightweight ways~\cite{Lee_NDSS2015}, which can also be used in our attack to pre-select the services to be targeted.

\smallskip\noindent\textbf{Mobile Platform.}
A non-negligible portion of users today access social web services from their mobile devices, so whether or not the attack is feasible in this realm is an important question to explore. For recent mobile platforms such as Android and iOS, the mechanics of most web browsers as well as the effective performance of the hardware and network are not significantly different from those of a PC; therefore, they are expected to yield sufficient RTT differences making our attack feasible. We partially proved this in our experiment with the tethering environment. The primary concern instead is the unique software ecosystem of mobile devices: many services encourage users to use a service-dedicated app instead of a browser to access their service. Even though some collaborative features such as social plug-ins or single sign-on may still urge some mobile users to log on via a browser, this ecosystem will surely limit the target coverage of our attack to a certain degree. We believe that a possible attack vector for this scenario which may need an attention might be an exploitation of a mobile platform-specific side channel, e.g.,
Android's Intent
and shared memory~\cite{chen2014peeking}, to bypass the app sandbox, analogical to how our attack exploited a browser timing side channel to bypass the same-origin policy, but we leave further discussions on this for a future study.

\smallskip\noindent\textbf{Limits on Blocking.}
For most services, limitations on the total number of users allowed to be blocked or the rate at which blocking requests can be issued from a single account are not explicitly stated. We have experimentally confirmed that at least ten million users on Twitter and three million users on Facebook and Tumblr were actually blockable over five days using a single account, and only DeviantArt and eBay seems to have had a limit on the maximum number of blocks per account. Also, Instagram appears to have had a limitation on the rate, i.e, the number of accounts that can be blocked per minute. 
As we have shown in Section~\ref{sec:userblocking}, neither disabling blocking nor posing a limit on it, is desirable from the viewpoints of the actual usage of the service and users' expectations. However, having limits on the total number of users to be blocked blocking may interfere with the process of building a high-coverage signaling account. Still, user-space partitioning would help alleviate this limitation and much of the effort for building signaling accounts is required just once, implying that attackers are not so exceedingly time-constrained when performing this task.

\smallskip\noindent\textbf{Length of Visit.}
As shown in Section~\ref{sec:experiments}, the attack can be executed in a realistically short time. In certain circumstances, however, such as when the RTT is high or when there is a need to use user-space partitioning, which increases the number of requests, it may be difficult to keep the user on the same webpage long enough for the JavaScipt code to finish. Even if the attack duration is short, because the behavior of a user is often unpredictable, a shorter attack is always preferable. A trivial approach to this problem is to prepare webpage content that is sufficiently ``attractive'' to cause the users to stay longer, but this is very user specific. Another solution is to save and restore the attack state between multiple attack sessions. By having the JavaScript code send partial results to the server as it attacks, even if the attack terminates before finishing, the attack can be resumed at another session from where it left off. Training data may be reused or not depending on the ``distance" between each attack session, e.g., the time elapsed between sessions.
Another solution is to open pop-up windows in the background or a tab and execute the attack there, hoping that the user would not notice or care to close it immediately.

\subsection{Ethics}
In Section~\ref{sec:userblocking}, all the data have been collected with user consent, and we followed guidelines presented by the ethics committee of Waseda University.

To evaluate the feasibility and impact of the attack techniques on social web service users, experimenting with attacks on actual social web services cannot be avoided.
All attacks in our experiment were checked manually and only generated a restricted amount of request.
As a result, our experiment was carefully controlled and only generated a restricted amount of traffic (requests), which did not increase the workload of the sites and did not undermine the quality of their services. 
Furthermore, our experiment performed against our own accounts.
Therefore, actual users of the services we examined were not directly involved in our attacks.

Even though the attack technique in this paper does not arise from a specific social web service, according to the principle of responsible disclosure, we have reported the details of our attacks and the experimental results to the relevant social web service providers and browser vendors to mitigate the attacks and improve future security design of social web. Several service providers and browser vendors have already finished implementing defenses and some are in the process.

\section{Defense}
\label{sec:defense}

In this section, we discuss defensive measures that can be taken against our attack. We emphasize that all approaches we are currently aware of either cause a serious degradation in the service quality or require considerable amounts of time and effort before being implemented or widely adopted. Nevertheless, we believe that raising the bar for an attacker would still be beneficial to the public.

\subsection{Server-side Defenses}

\smallskip\noindent\textbf{Token Validation.} 
Token-based defenses are widely adopted to prevent CSRF attacks in general. The server appends a one-time random string, or token, to each URL link generated and verifies it when the link is accessed. This prevents any third-party from generating a valid link; therefore, the attacker will not be able to receive valid responses containing information useful for the attack as long as the token-checking process is applied before the block checking at the server side. A major drawback of this defense is that legitimate requests are also affected and result in consequences such as breaking search engine results or prohibiting any means of link sharing, including those on blog posts and emails.
Promising approaches which acquire user contents by using JavaScript's XMLHttpRequest with a valid token such as {\it placeholder}~\cite{Goethem_CCS2015} and {\it double-submit cookie}~\cite{OWASP_CSRF} have been proposed, but they still require a change in the system architecture design and also the delay caused by the extra hop may negatively affect the user experience.

\smallskip\noindent\textbf{Response Time Control.} 
The server could adjust the response time to minimize the block/non-block RTT difference. One approach is to artificially equalize the response times by adding delays to whichever has the shorter response time. Another approach is to randomize the response time by injecting delays of random lengths. However, either approach would impose a non-negligible performance degradation experienced by the user. In general, this type of timing side-channel defense is difficult to perfect; the profound study results in this area provide advanced attackers with various ways to amplify such differences at the cost of some increased effort, as we also have exemplified in this paper. In addition, the network delay is often uncontrollable from the service side so a perfect control is difficult to attain from the server side. Note that such types of server-side defenses are often thwarted by other timing side-channel approaches, such as those leveraging the content cache~\cite{Goethem_CCS2015}.

\smallskip\noindent\textbf{Usage Restriction.}
Our attack, when implemented in a straightforward manner, may exhibit behavioral characteristics not usually seen in the normal usage of the service. One case of such an anomaly would occur in the preparation process of a signaling account, which requires a massive number of blocking requests to be issued within a short time. Another is in the process of launching the attack from a browser, which causes an abnormal number of GET requests to be issued. The service can either restrict this in the form of the rate limit, CAPTCHA, or some means of heuristic anomaly detection. However, these defenses are expected to function only as a mild mitigation, because advanced attackers have historically been able to circumvent these types of defenses. The most extreme form of restriction is to remove the user-blocking capability from the service. All these types of restriction-based measures, however, lead to an undermining of the ability to suppress those who truly needs to be blocked, which may result in a degradation of the service quality.

\subsection{Client-side Defenses}

\smallskip\noindent\textbf{User.}
Defenses that can be taken by a user alone are limited to quite trivial ones. One approach is to isolate the browsing environment in which the web service is used, from that used for other purposes. This can be done, for example, by using the private browsing feature commonly available in modern browsers, logging out of the service when not in use, or simply using a different browser. Another approach is to restrict the execution of JavaScript using browser plug-ins such as NoScript\cite{noscript}, which would severely impair the attacker's capability to carry out such an attack. Obviously, all of these measures greatly increase the user's cost of not only using the service but also web browsing in general. Further, it would deactivate some features such as social plug-ins or advertisements that benefit both of the user and the service provider.

\par\smallskip\noindent\textbf{Web Browser.}
{\it SameSite}~\cite{ssc} is a cookie attribute that allows flexible control of sending cookies in cross-site requests.
This interferes with the functionality of some social plug-ins, but otherwise it is a very elegant solution.
To use this feature, it is necessary for the browser adopting it, and the web service explicitly declare it in the HTTP header.

Equalizing the response times, for example, by injecting delays to the processing time, is also a possible measure that can be taken on the browser side. Further, the detection of anomalies such as frequent errors resulting from failed rendering may be another option. However, these approaches are often only viable for a certain class of timing side channels; they tend to be thwarted eventually by other newly developed timing attacks using different approaches, as exemplified by the attack using the browser cache mentioned in another study~\cite{Goethem_CCS2015}. 

\section{Related Work}
\label{sec:related}
We present previous studies concerning timing attacks, which is the fundamental technique of our method uses to compromise user's privacy.
In addition, we introduce other side-channel leaks based on the browser functionality and methods to identify and track users.

\subsection{Web-based Timing Attacks}
\label{related_timingattack}
A timing attack is one type of side-channel attack that has been studied primarily in cryptography for more than two decades.
It typically exploits the execution time or power consumption of a cryptosystem to infer secret key and private information~\cite{Kocher_CRYPTO1996,Kocher_CRYPTO1999}.
Studies of timing attacks have expanded to web-based systems regardless of the cryptosystem that exploits the communication time and size of the web content.
Bortz et al. presented a pioneer work on web-based timing attacks; they classified web-based timing attacks into {\it direct timing} and {\it cross-site timing}~\cite{Bortz_WWW2007}.
Our proposed method is classified as a web-based cross-site timing attack.

A direct timing attack directly measures the response times from a system, e.g., a website, to extract private information from a system.
Bortz et al. proposed a method to expose valid user names and the number of private photos from a website by measuring the response time of HTTP~\cite{Bortz_WWW2007}.

Cross-site timing attacks indirectly measure the response times or content size of web on a browser to extract private information from a browser or website.
It enables a malicious website to obtain information about the target browser's view of another website using cross-site content that often violates the same-origin policy~\cite{SameOriginPolicy}.
Methods to break the same-origin policy and their countermeasures have been presented since 2000~\cite{CERT_XSS,Kirda_SAC2006,Vogt_NDSS2007,Stock_Security2014}; however, the many of cross-origin techniques are still effective on modern web browsers.
Liang et al. leveraged several CSS features to indirectly monitor the rendering of a target resource~\cite{Liang_DSN2014}.
Goethem et al. proposed a cache-based timing attack using HTML5 functionalities, which can bypass the same-origin policy, to estimate the size of a cross-origin resource~\cite{Goethem_CCS2015}.
Gelernter et al. presented a {\it cross-site search attack} on well-known web services to distinguish between the loading time of empty and full responses, which enables an attacker to distinguish sensitive data of target users in the records of the web services~\cite{Gelernter_CCS2015}.
Jia et al. demonstrated a {\it geo-location inference attack} on well-known web services, by using the load time of location-sensitive resources left by geography-specific websites (e.g., Google's local domain)~\cite{Jia_IC2015}.
Our method is not new in the context of cross-site attacks; however, the idea is unique in that user blocking, which is a fundamental functionality of social webs, can be used to distinguish between blocked and, consequently, to identify their social accounts.

\subsection{Side-channel Leaks on Browsers}
\label{related_sidechannel}
A side-channel attack on a browser without timing features is another class of privacy attack.
To infer the status of a cross-origin resource, Lee et al. developed a {\it URL status identification attack} using {\tt ApplicationCache} that exploits cross-origin resource caching~\cite{Lee_NDSS2015} and they suggested advanced privacy threats using this attack, e.g., login status determination and internal web server probing.
A {\it history-stealing attack} is a typical attack that extracts the browsing history of URLs~\cite{Jacson_WWW2006,Wondracek_SP2010}.
This attack depends on the fact that a web browser handles CSS properties of URL hyperlinks differently depending on whether the URL was previously accessed by the web browser~\cite{Ruderman_CSSattack}, which leads to allowing a client-side script to access such properties.
To fix this, Baron proposed a solution that blocks scripts from accessing the CSS properties of hyperlinks, and all popular browsers (e.g., Firefox, Chrome, Safari, and IE) have adopted this solution. 
As a result, this type of history stealing attack no longer works in the latest versions of these browsers~\cite{Baron_historystealing,W3C_historystealing}.

\subsection{Social Account Identification}
While various methods have been proposed to effectively track browsers on the Internet (e.g., cookies, browser cache, and browser fingerprints~\cite{Eckersley_panopticlick,Roesner_NSDI2012,Acar_CCS2014}), 
these tracking methods focus on identifying distinct browsers rather than the user of the browsers.
The goal of our proposed method is to identify the user (i.e., the social account) which differs from the above browser tracking methods.
Many of the studies introduced in Sections~\ref{related_timingattack} and \ref{related_sidechannel} mentioned that their proposed methods could be used for inferring the status of social account or identifying social account~\cite{Bortz_WWW2007,Lee_NDSS2015,Wondracek_SP2010}.
The difference of response time of login page was used for inferring account validity~\cite{Bortz_WWW2007}. 
With a similar motivation, conditional redirections of the HTTP URLs was used for distinguishing whether a victim web browser is logged in to the web service~\cite{Lee_NDSS2015}.
The combination of {\it group membership information}, e.g., group ID or group directory in browser's access history, was used for identifying a social account~\cite{Wondracek_SP2010}.
These differences are extracted from previously provided pages, e.g., login pages and group membership pages.
In contrast, our method is unique in that an attacker can fully control the visibility of pages in order to create discriminable differences. 

\section{Conclusion}
\label{sec:conclusion}
This work presents a practical side-channel attack that identifies the social account of a user visiting the attacker's website. It exploits the user-blocking mechanism, or the visibility control property, commonly available in most social web services today to create a controllable side channel that provides the attacker with complete and flexible control over the leaked information, be it informative enough to uniquely identify the user or be it highly resilient to noise. With experiments, we demonstrated that our attack is in fact applicable to current mainstream social web services today and we argued that defending against this threat would not be easy without imposing a negative impact on the relevant services. It is ironic that the blocking feature designed to suppress harmful users can now be turned against harmless users; some form of mitigation is urgent and a reworking of the design of this feature is suggested as a future work. 

\bibliographystyle{IEEEtran}
\bibliography{ref}

\end{document}